\documentstyle[12pt]{article}

\newcommand{\be}{\begin{equation}}
\newcommand{\ee}{\end{equation}}
\newcommand{\bea}{\begin{eqnarray}}
\newcommand{\eea}{\end{eqnarray}}

\newcommand{\GX} {\gamma_{xx}}
\newcommand{\GT} {\gamma_{tt}}
\newcommand{\GG} {\gamma_{tx}}
\newcommand{\GD} {\gamma_{\eta\eta}}
\newcommand{\DE} {\delta_{\epsilon}}
\newcommand{\DW} {\delta_{\omega}}

\begin{document}

\begin{center}
\begin{large}
{\bf  Holographic Central Charge \\}
{\bf  for \\}
{\bf  2-Dimensional de Sitter Space  \\}
\end{large}  
\end{center}
\vspace*{0.50cm}
\begin{center}
{\sl by\\}
\vspace*{1.00cm}
{\bf A.J.M. Medved\\}
\vspace*{1.00cm}
{\sl
Department of Physics and Theoretical Physics Institute\\
University of Alberta\\
Edmonton, Canada T6G-2J1\\
{[e-mail: amedved@phys.ualberta.ca]}}\\
\end{center}
\bigskip\noindent
\begin{center}
\begin{large}
{\bf
ABSTRACT
}
\end{large}
\end{center}
\vspace*{0.50cm}
\par
\noindent

Recently, investigations have begun into a holographic duality
for two-dimensional de Sitter space. In this paper, we
evaluate the associated central charge, using a modified version
of the canonical Hamiltonian method that was first advocated
by Catelani {\it et al}. Our computation agrees with
that of  a prior work  (Cadoni {\it et al}), but we argue that the
method used here is, perhaps, aesthetically preferable on holographic
grounds. We also confirm an agreement between the Cardy
and thermodynamic entropy, thus providing further
support for the conjectured two-dimensional
de Sitter/conformal field theory correspondence.

\newpage

\par

\section{Introduction}

To this date, the greatest triumph of the holographic principle
\cite{THO,SUS} remains the  correspondence
between anti-de Sitter (AdS) bulk spacetimes
and conformal field theories (CFTs) of one dimension fewer 
\cite{MAL,GUB,WIT}. 
On the basis of this success, there has been substantial
interest in establishing a similar duality
in a de Sitter (dS) context. There has indeed been
undeniable  progress in this direction \cite{STR2,WIT2}
(and, for instance, \cite{KLE}-\cite{BAL}), although
unresolved issues still remain. (For a recent critique, see
\cite{SUS2}.)
\par
For the most part, investigations into the (conjectured)
dS/CFT correspondence have concentrated on three, four and five
bulk dimensions; with three being the simplest case,
and four and five being (presumably) the physically
most relevant. Nonetheless, the intriguing case of two
bulk dimensions has begun to draw some limited
attention \cite{NS,CAD2}. These works (\cite{CAD2} in
particular) have been significantly inspired
by prior treatments on the AdS$_2$/CFT$_1$ correspondence
(including  \cite{STR4}-\cite{CAD5}).
\par
Naively, one might expect that holographic dualities
in a two-dimensional bulk context would be the simplest
cases of all. This may certainly be true
on a calculational level; however, one finds 
 such two-dimensional
dualities to be plagued by conceptually ambiguous features. 
For instance, from a bulk perspective,  it
is  conspicuously unclear as to how one should formulate
a  two-dimensional theory of gravity.
This ambiguity is a direct consequence of the
``conventional'' Einstein
tensor identically vanishing in two-dimensions of spacetime.
The standard method of circumvention
is to introduce another dynamical field into the mix;
typically, an auxiliary scalar field known as the dilaton.
However, there is generally no {\it a priori} rationale
for modifying an otherwise pure gravitational theory
in this manner.\footnote{A notable exception is when the
two-dimensional theory is regarded as having a higher-dimensional
pedigree; in which case, the dilaton is usually associated 
with a compactified dimension.} Furthermore, from
the viewpoint of the conformal boundary theory, it is 
somewhat problematic that  there is, formally speaking,
no such entity as a one-dimensional conformal field theory.
Hence, any notion of a straightforward (A)dS$_n$/CFT$_{n-1}$ 
correspondence is sabotaged when $n=2$. However, this last
point is not quite as bleak as it sounds: a ``CFT$_1$''
can be feasibly interpreted as (for instance)
a discrete light-cone quantization of a CFT$_2$ \cite{STR4}
or a conformal-mechanical system coupled to an external
source \cite{CAD7}.
\par
Let us now refocus our attention on the recent investigation
into the dS$_2$/CFT$_1$ duality  by Cadoni {\it et al} \cite{CAD2}.
One of the highlights of this work was a calculation of the central
charge associated with a holographically induced Virasoro
algebra. (Significantly, the Virasoro algebra effectively describes
the two-dimensional conformal group \cite{FMS}.) To this end,
the authors employed a canonical Hamiltonian framework \cite{DIR,HAM}.
More specifically, they deduced the central charge by considering the
deformation algebra of the  symmetry generators
associated with the Hamiltonian asymptotic-boundary terms \cite{BH}.
One technical caveat however: to obtain
the ``correct''  the central charge (i.e., the value
for which the Cardy formula \cite{CAR} agrees with the
thermodynamic entropy), it was also necessary to include
the contribution from an inner boundary of the spacetime.
(Notably, the  same adjustment was used in the analogous AdS$_2$/CFT$_1$
calculation \cite{CAD7}.) Although this is a technically sound procedure,
it is arguably in conflict with the very essence
of the holographic principle. That is to say, one might
expect that a {\it single} connected boundary (such as
a ``preferred screen'' \cite{BOU}) would be 
capable of encoding all of the relevant information about the
bulk spacetime. If this were the case, the asymptotic boundary
theory should be, by itself,  sufficient to
reproduce the correct value of the central charge.
\par
With the above discussion in mind, let us now consider a
prior work, specific to the AdS$_2$/CFT$_1$ correspondence,
by Catelani, Vanzo and Caldarelli (CVC) \cite{VAN}.
These authors advocated a ``modified'' canonical-Hamiltonian treatment
as a means for resolving an apparent discrepancy between
the statistical  Cardy entropy   and the thermodynamic entropy.
(Historically speaking, the two entropies differed by a factor
of $\sqrt{2}$ in the original AdS$_2$/CFT$_1$ calculation \cite{CAD3}.
This discrepancy was later resolved by several distinct means:
 sacrificing diffeomorphism invariance \cite{NN1}, the forementioned
 inclusion of  the inner-boundary contribution \cite{CAD7},
and the method currently under consideration.) More to the
point, CVC have argued that one should rigorously
keep track of the various boundary terms that arise while
formulating the canonical Hamiltonian. Note that this philosophy
is contrary to the conventional practice of ignoring the boundary terms
until they are explicitly needed. (In principle, one can retrieve
the discarded terms by enforcing that  the bulk Hamiltonian
is differentiable. See, for instance, \cite{LGK}.) 
\par
As it turned out, the modifications suggested by CVC did
indeed lead to a precise agreement between the statistical
Cardy entropy and the AdS$_2$ thermodynamic entropy.
Moreover and significantly to our prior discussion, the central charge,
as calculated by the CVC method, is strictly accounted for
 by the symmetry properties of a single (asymptotic) boundary.
The purpose of the current paper is to see if the same success 
can be achieved in a two-dimensional de Sitter  context.
\par
The remainder of the paper is organized as follows.
In Section 2, we introduce  the relevant  dS$_2$ formalism,
with the focus being on asymptotic symmetries.
In Section 3, by way of a  ``modified'' Hamiltonian framework
(as discussed above), we calculate the central charge
of the dually related CFT.  Section 4 considers
the Cardy  statistical entropy.  Finally,
a brief summary is provided in Section 5.

\section{Preliminary Analysis}

In considering a two-dimensional theory of 
gravity,  it is convenient to work with
 an explicit form for the action.
Here, the most natural choice is the de Sitter analogue
of Jackiw-Teitelboim gravity \cite{JT}. That is:
\be
I={1\over 2}\int d^2x\sqrt{-g}\eta \left[R-{2\over l^2}\right],
\label{1}
\ee
where $\eta$ is the dilaton (scalar) field and $l$ is a
fundamental parameter of dimension length.  Let us take note
of the field equations, which are obtained by varying the action with
respect to the dilaton and metric (respectively):
\be
R={2\over l^2},
\label{2}
\ee
\be
\Box \eta= -{2\over l^2}\eta.
\label{3}
\ee
The former  clearly indicates the desired property of
a metric with a constant, positive curvature, whereas the latter usefully
describes the dynamics of the dilaton.
\par
To obtain the general, classical solution for the action (\ref{1}),
we can begin with the well-known Jackiw-Teitelboim solution
\cite{JT}:
\be
ds^2_{JT}=-\left({r^2\over l^2}-a^2\right)d\tau^2+
\left({r^2\over l^2}-a^2\right)^{-1}dr^2,
\label{4}
\ee
\be
\eta_{JT}=\eta_0 {r\over l},
\label{5}
\ee
(where $a$ and $\eta_0$ are constants of integration)
and then analytically continue from $l^2$ to $-l^2$.
Along with an appropriate relabeling of the coordinates,
this process yields:
\be
ds^2=-\left({t^2\over l^2}+a^2\right)^{-1}dt^2+
\left({t^2\over l^2}+a^2\right)dx^2,
\label{6}
\ee
\be
\eta= \eta_0 {t\over l}.
\label{7}
\ee
Note that we have also rescaled the dilaton field, which
is permissible via the arbitrariness of $\eta_0$.
(Also note that we will, without loss of generality,
regard $\eta_0$ to be positive.)
\par
At this point, some commentary on global properties is in order.
In a strictly technical sense, the above solution
covers the entire spacetime manifold ($-\infty\leq t$,$x\leq +\infty$).
However, it can be argued (for instance, \cite{CAD1}) that
the surface of vanishing dilaton (in our case, at $t=0$) should
act as a boundary of the spacetime. This argument follows from the
observation that $\eta^{-1}$ effectively serves as the gravitational
coupling ``constant'' in two-dimensional gravity theories.
On this basis, we will view the manifold as being limited
to non-negative values of $t$. Note that the asymptotic boundary 
at $t\rightarrow\infty$ complies with the usual (de Sitter)
notion of spacelike future infinity or ${\cal I}^{+}$.\footnote{Conversely,
we could have, just as legitimately, limited the manifold
to non-positive values of $t$; in which case, the asymptotic boundary
would describe spacelike past infinity or ${\cal I}^{-}$.}
\par
Although the spacetime is lacking  a conventional event
horizon (or any black hole-like structure),
one can still anticipate thermodynamic properties in analogy
to those associated with a black hole. This presumption
follows from the existence of a surface, at $t=0$, behind
which nothing can communicate with infinity. That is to say,
such a surface naturally leads to an entanglement-like
entropy and its associated  thermodynamic structure.
As in higher-dimensional de Sitter spacetimes, the absence
of a globally time-like Killing vector complicates
any attempt at rigorously calculating the thermodynamics
in question.\footnote{Although not fatally so. See, for instance,
\cite{BDM}.}  Nonetheless, one can still anticipate
that the various thermodynamic properties are, at a classical level,
essentially equivalent to those of a Jackiw-Teitelboim
black hole. This follows from the two spacetimes being related 
by an almost trivial analytic continuation.\footnote{Such an argument
has, of course, no validity when quantum corrections
become important.} Indeed, some quantitative analysis of
the action (\ref{1}) has already come out in support of
this conjecture \cite{CAD2}. Thus,  we can utilize the following 
thermodynamic identities (as measured by an observer at infinity)
\cite{GKL}:
\be
T={|a|\over 2\pi l},
\label{9}
\ee
\be
|M|={a^2 \eta_0\over 2 l},
\label{8}
\ee
\be
S=2\pi\eta_o |a|,
\label{10}
\ee
with these quantities representing the classical values of temperature,
conserved mass and entropy (respectively). Note the ambiguity
in the sign of the conserved mass, which is an inherent feature of
de Sitter space regardless of the  dimensionality (see the ``note added''
in \cite{KLE}).
\par
Let us now extend considerations to spacetimes that are,
in some sense, merely ``asymptotically de Sitter''. Closely following
prior studies on (A)dS$_3$ and (A)dS$_2$ (most recently, \cite{CAD2}),
we will say that a solution is asymptotically de Sitter
if it conforms to the following set of  boundary conditions 
as $t\rightarrow\infty$:
\be
g_{xx}={t^2\over l^2}+\GX(x) +{\cal O}(t^{-2}),
\label{11}
\ee
\be
g_{tt}=-{l^2\over t^2}+{l^4\over t^4}\GT(x) +{\cal O}(t^{-6}),
\label{12}
\ee
\be
g_{tx}={l^3\over t^3}\GG(x)+{\cal O}(t^{-5}),
\label{13}
\ee
\be
\eta=\eta_0\left[{t\over l}\rho(x)+{l\over t}{\GD(x)\over 2}+
{\cal O}(t^{-3})\right],
\label{14}
\ee
where the $\gamma$'s and $\rho$ are model-dependent functions.
Note that $\rho\sim 1$ near the classical configuration.
\par
The most general diffeomorphisms which preserve
the above boundary conditions are found to be as follows:\footnote{Here
and throughout, a prime (dot) denotes a spatial (temporal)
differentiation.} 
\be
\xi^{x}[\epsilon]=\epsilon(x)+{1\over 2}{l^4\over t^2}\epsilon^{\prime\prime}
(x) +{\cal O}(t^{-4}),
\label{15}
\ee
\be
\xi^{t}[\epsilon]=-t\epsilon^{\prime}(x) +{\cal O}(t^{-1}),
\label{16}
\ee
where $\epsilon$ is an arbitrary function of $x$.
Note that $\delta_{\epsilon}\eta={\cal L}_{\epsilon}\eta
\sim {\cal O}(t)\sim {\cal O}(\eta)$, which 
may seem severe, but is a necessary consequence of  preserving
the scalar nature of the dilaton (i.e., preserving  diffeomorphism
invariance of the action) \cite{CAD3}. Further note that,
when $\epsilon =0$, the higher-order terms can be identified
with  pure gauge transformations \cite{STR}.
\par
It is straightforward to show that, up to gauge transformations,
the boundary fields of Eqs.(\ref{11}-\ref{14}) transform
as follows:
\be
\DE\GT=\epsilon\GT^{\prime}+2\epsilon^{\prime}\GT,
\label{17}
\ee
\be
\DE\GX=\epsilon\GX^{\prime}+2\epsilon^{\prime}\GX+
l^2\epsilon^{\prime\prime\prime},
\label{18}
\ee
\be
\DE\GG=\epsilon\GG^{\prime}+3\epsilon^{\prime}\GG
-l\epsilon^{\prime\prime}\left(\GT+\GX\right),
\label{19}
\ee
\be
\DE\rho=\epsilon\rho^{\prime}-\epsilon^{\prime}\rho,
\label{20}
\ee
\be
\DE\GD=\epsilon\GD^{\prime}+\epsilon^{\prime}\GD
+ l^2 \epsilon^{\prime\prime}\rho^{\prime}.
\label{21}
\ee
\par
It is a further point of interest that, if one sets $\epsilon=0$,
and thus considers pure gauge transformations:
\be
\xi^{x}[0]={l^4\over  t^4}\alpha^x(x),
\label{22}
\ee
\be
\xi^{t}[0]={l \over t}\alpha^t(x),
\label{23}
\ee
than the theory is found to have (up to constant factors) only
two gauge invariant quantities. These being $\rho(x)$ and:
\be
\theta(x)\propto \GX-{1\over 2}\GT.
\label{24}
\ee
As a consequence, we are always free to choose a gauge
for which $\GG=0$. This luxury will prove to be convenient
in later analysis.
\par
Finally, it is worth noting that the dilatonic field
equation (\ref{3}), when translated to the solution of
Eqs.(\ref{11}-\ref{14}), yields the relation:
\be
l^2\rho^{\prime\prime}=\rho(\GT-\GX)-\GD.
\label{25}
\ee
This relation will prove to be another useful
tool later on. 

\section{Hamiltonian Formalism}

A canonical Hamiltonian framework \cite{DIR,HAM} has often
been used, with considerable success, 
for studies on two-dimensional dilatonic gravity (for instance,
\cite{LGK}). In such studies, the prevailing wisdom
has been that the  various boundary terms (which arise whenever
one differentiates by parts) can be harmlessly ignored
until later in the calculation.
 Significantly to  
this approach, one can (in principle)  ultimately retrieve the discarded 
surface terms 
by imposing   differentiability on the bulk Hamiltonian.
On the other hand, Catelani {\it et al} have recently argued \cite{VAN}
that,  to calculate the central charge
of the conformal boundary theory, one should pedantically keep
track of all boundary terms throughout the process.
In the analysis to follow, we will endeavor to extend this latter
philosophy to the dS$_{2}$ model of the prior section.
\par
Let us begin here by employing an ADM-like decomposition of the 
metric, as appropriate for a spacelike boundary:
\be
ds^2=N^2dx^2-\sigma^2\left(dt+Vdx\right)^2,
\label{26}
\ee
where $N$, $\sigma$ and $V$ are 
gauge-dependent functions of the spacetime coordinates.
With this parametrization, the action of Eq.(\ref{1})
takes the form:
\be
I=I_{bulk}+I_{\infty},
\label{27}
\ee
where:
\be
I_{bulk}=\int dtdx \left[{\eta^{\prime}\over N}\left(\sigma^{\prime}
-{\dot V}\sigma-V{\dot\sigma}\right)
+{{\dot\eta}V\over N}\left({\dot V}\sigma+V{\dot\sigma}-\sigma^{\prime}
\right)-{{\dot\eta}{\dot N}\over \sigma}-{\eta\over l^2} N \sigma
\right]
\label{28}
\ee
\be
I_{\infty}=\int dx \left[{\eta{\dot N}\over \sigma}
+{\eta V\over N}\left(\sigma^{\prime}-{\dot V}\sigma
-V{\dot \sigma}\right)\right]_{{\cal I}^{+}}.
\label{29}
\ee
Note that we have, in the prescribed manner,
 explicitly included any surface term 
arising at spacelike future  infinity (${\cal I}^{+}$).
Strictly speaking, there are also surface 
terms associated with the inner boundary, and
these will play an important role in constraining
the  geometry near $t=0$. 
 However, such terms should  have no influence on 
the action of the asymptotic conformal  group \cite{VAN}
 and, hence,  will be left out of the current discussion.
\par
Varying the total action (\ref{27}) with respect
to small deformations in the four fields
($\eta$, $\sigma$, $N$, $V$), we obtain a series of variational
terms at ${\cal I}^{+}$.  We are able to eliminate
many of these terms by sensibly imposing that
the geometry is fixed at the asymptotic boundary;
that is, $\delta \eta =\delta\sigma=\delta N=\delta V=0$
at ${\cal I}^+$.  After which, the following terms still remain:
\be
\int dx \left[{\eta\over \sigma}\delta{\dot  N}
+{\eta V\over N}\left(\delta\sigma^{\prime}-\sigma\delta{\dot  V}
-V\delta{\dot \sigma}\right)\right]_{{\cal I}^{+}}.
\label{30}
\ee
\par
It is now evident that, for a well-defined variational
principle, the action should be supplemented by a surface contribution
whose variation precisely cancels Eq.(\ref{30}).
Denoting this supplementary term as ${\tilde I}_{\infty}$, we
have:
\be
{\tilde I}_{\infty}=-\int dx \left[{\eta{\dot N}\over \sigma}
+{\eta V\over N}\left(\sigma^{\prime}-{\dot V}\sigma
-V{\dot \sigma}\right)\right]_{{\cal I}^{+}}.
\label{31}
\ee 
\par
Ignoring the boundary supplement (${\tilde I}_{\infty}$)
for just a moment, we know that the original action 
(\ref{27}) gives rise to the following canonical
Hamiltonian up to surface terms:\footnote{Here,
we are following the standard Dirac program \cite{DIR},
except that time and position have exchanged their
usual roles.}
\be
H_{bulk}=\int dt\left[\Pi_{\eta}\eta^{\prime}
+\Pi_{\sigma}\sigma^{\prime}
+\Pi_{N}N^{\prime}
+\Pi_{V}V^{\prime}-{\cal L}\right],
\label{32}
\ee
where  the Lagrangian (${\cal L}$) and canonical momenta
($\Pi_{\psi}$) will be defined in the usual manner. That is:
\be
\int d^2x {\cal L}= I_{bulk},
\label{33}
\ee
\be
\Pi_{\eta}={\delta {\cal L}\over \delta \eta^{\prime}}
={1\over N}\left[\sigma^{\prime}-{\dot V}\sigma-
V{\dot \sigma}\right],
\label{34}
\ee
\be
\Pi_{\sigma}={\delta {\cal L}\over \delta \sigma^{\prime}}
={1\over N}\left[\eta^{\prime}-{\dot \eta}V \right],
\label{35}
\ee
\be
\Pi_{N}={\delta {\cal L}\over \delta N^{\prime}}
= 0,
\label{36}
\ee
\be
\Pi_{V}={\delta {\cal L}\over \delta V^{\prime}}
=0.
\label{37}
\ee
\par
Some straightforward calculation verifies that 
$H_{bulk}$
can be expressed  as a linear
combination of  ``weakly vanishing'' (in a Dirac sense \cite{DIR}) constraints.
In particular, one finds the following up to surface terms: 
\be
H_{bulk}=\int dt\left[
V {\cal H}_V +N{\cal H}_N \right],
\label{38}
\ee
where:
\be
{\cal H}_V ={\dot \eta}\Pi_{\eta}-\sigma{\dot\Pi}_{\sigma}\approx 0,
\label{39}
\ee
\be
{\cal H}_N =\Pi_{\eta}\Pi_{\sigma}-{{\ddot\eta}\over\sigma}
+{{\dot\eta}{\dot\sigma}\over\sigma^2}+{1\over l^2}\eta\sigma
\approx 0.
\label{40}
\ee
Note,
however, that the transformation from   
Eq.(\ref{32}) to Eq.(\ref{38}) gives rise to the following
asymptotic surface terms (which we collectively
denote as ${\tilde H}_{\infty}$):
\be
{\tilde H}_{\infty}=-\left[\eta^{\prime}{V\over N}\sigma+ {\dot\eta}
\left({N\over\sigma}-{V^2\sigma\over N}\right)\right]_{{\cal I}^+}.
\label{41}
\ee
\par
Let us re-emphasize that the bulk Hamiltonian has
been formulated to vanish on the constraint surface.
At this point,
one normally considers (for instance, \cite{LGK}) 
 a total Hamiltonian of the form:
$H=H_{bulk}+J$, where $J$ is a surface contribution that
ensures the differentiability of $H$ (i.e., ensures $\delta H=0$
 on the relevant boundaries). Alternatively, we
will follow Catelani {\it et al} \cite{VAN} and consider:
\be
H=H_{bulk}+J+K,
\label{41.5}
\ee
where $J$ is still defined as  above; whereas $K$
is a  surface contribution that accounts for both 
${\tilde H}_{\infty}$ (\ref{41}) and  the  ``supplementary'' action
term 
${\tilde I}_{\infty}$ (\ref{31}). 
More precisely, let us
first define
$\int dx {\tilde{\cal L}}_{\infty}\equiv {\tilde I}_{\infty}$ and then:  
\bea
K&\equiv& {\tilde H}_{\infty}-{\tilde {\cal L}}_{\infty} \nonumber \\
&=& \left[{1\over\sigma}\left(\eta{\dot N}-{\dot\eta}N\right)
+V\left(\eta\Pi_{\eta}-\sigma\Pi_{\sigma}\right)\right]_{{\cal I}^+},
\label{42}
\eea
where Eqs.(\ref{34},\ref{35}) have been  used to simplify the
lower line.
\par
Let us now see what  we can learn about asymptotic symmetries
in this Hamiltonian framework.
Inasmuch as  $H_{bulk}\approx 0$,
it is, for our purposes, sufficient to consider 
the action  of the surface terms, $J$ and $K$.
Beginning with $J$ and employing the above prescription,
we have:
\bea
\delta J&=& -\left.\delta H_{bulk}\right|_{{\cal I}^+}
\nonumber \\ &=&
\left[{N\over\sigma}\left(\delta{\dot\eta}-
{{\dot\eta}\over\sigma}\delta\sigma\right) -{{\dot N}\over\sigma}
\delta\eta+V\left(\sigma\delta\Pi_{\sigma}-\Pi_{\eta}\delta\eta\right)
\right]_{{\cal I}^+}.
\label{43}
\eea
\par
An underlying premise of the canonical Hamiltonian framework is that
a given symmetry generator, 
$\xi^{\mu}[\epsilon]$, has an associated conserved charge, 
$J[\epsilon]$, which can be expressed as (for instance, \cite{CAD3}):
\be
\delta J[\epsilon]=\left[{\epsilon^{\perp}\over\sigma}\left(\delta{\dot\eta}-
{{\dot\eta}\over\sigma}\delta\sigma\right) 
-{{\dot \epsilon}^{\perp}\over\sigma}
\delta\eta+\epsilon^{||}
\left(\sigma\delta\Pi_{\sigma}-\Pi_{\eta}\delta\eta\right)
\right]_{{\cal I}^+},
\label{44}
\ee
where $\epsilon^{\perp}=N\epsilon^x$ and
$\epsilon^{||}=\epsilon^t + V\epsilon^x$.
\par
For the purpose of evaluating Eq.(\ref{44}), let
us  now reconsider Eqs.(\ref{11}-\ref{14}) for
 the asymptotic form of the metric.
We can further fix the gauge, without loss of 
generality,\footnote{See the discussion
immediately following Eq.(\ref{24}).} so that  $\GG=0$.
Comparing with Eq.(\ref{26}),  we then
 obtain $V=0$ and the following relations:  
\be
N={t\over l}+{l\over t}{\GX(x)\over 2} +{\cal O}(t^{-3}),
\label{45}
\ee
\be
\sigma={l\over t}-{l^3\over t^3}{\GT(x)\over 2} +{\cal O}(t^{-5}),
\label{46}
\ee
\be
\eta=\eta_0\left[{t\over l}\rho(x)+{l\over t}{\GD(x)\over 2}\right]+
{\cal O}(t^{-3}).
\label{47}
\ee
\par
Directly incorporating the above formalism into Eq.(\ref{44}),
we find:
\be
\delta J[\epsilon]=\eta_0\left[{\epsilon\over l}\left(\GX\delta\rho
+{\rho\over 2}\delta\GT-\delta\GD\right)-l\epsilon^{\prime}
\delta\rho^{\prime}+l\epsilon^{\prime\prime}\delta\rho\right],
\label{48}
\ee
where only terms that are finite as $t\rightarrow\infty$
have been retained.
\par
Our next objective is to obtain a similar expression for
the remaining surface contribution, $K$ (\ref{42}). In analogous
fashion, the associated conserved charge, $K[\epsilon]$,
takes the form:
\be
K[\epsilon]=\left[{1\over\sigma}\left({\dot\epsilon}^{\perp}
\eta-\epsilon^{\perp}{\dot\eta}\right)+\epsilon^{||}
\left(\eta\Pi_{\eta}-\sigma\Pi_{\sigma}\right)
\right]_{{\cal I}^+}.
\label{49}
\ee
Reapplying Eqs.(\ref{45}-\ref{47}), the gauge choice $V=0$
and the
$t\rightarrow\infty$ limit, we have:
\be
K[\epsilon]=\eta_0\left[{\epsilon\over l}\left(\GD
-\rho\GX\right)+l\epsilon^{\prime}
\rho^{\prime}-l\epsilon^{\prime\prime}\rho\right].
\label{50}
\ee
\par
Let us now define a ``total''
conserved charge:
$\Psi[\epsilon]\equiv J[\epsilon]+K[\epsilon]$.
Summing Eq.(\ref{44}) and the variation of 
Eq.(\ref{50}), we then obtain:
\be
\delta \Psi[\epsilon]=\eta_0{\epsilon\rho\over 2l}
\left[\delta\GT-2\delta\GX\right].
\label{51}
\ee
It is a point of interest, and certainly no coincidence, that the
conserved charge can be expressed strictly in terms of
gauge invariant quantities: $\rho$ and $\theta$ of Eq.(\ref{24}).
\par 
Next, we will focus on configurations near the classical solution;
that is,
$\rho(x)=1+{\overline \rho}(x)$, where
${\overline \rho}(x)$ and the
$\gamma$'s are much less than 1. Some manipulation of Eq.(\ref{51})
and application of the perturbative field equation (\ref{25}) then
leads to:
\be
 \Psi[\epsilon]=l\eta_0\epsilon
{\overline \rho}^{\prime\prime}\pm\epsilon M.
\label{52}
\ee
Here, $M$ is an integration constant, which can be directly
identified, up to a sign ambiguity, with the conserved mass
of Eq.(\ref{8}). This identity follows from the observation
that, for a purely classical, on-shell solution, one
should obtain 
$H[\epsilon]=\epsilon |M|$.
\par
Applying the  transformation equations (\ref{17}-\ref{21})
to Eq.(\ref{52}) and defining  $\epsilon\Theta\equiv\Psi[\epsilon]$,
 we find:
\be
\DW\Theta=\omega\Theta^{\prime}+2\omega^{\prime}\Theta-
l\eta_0\omega^{\prime\prime\prime},
\label{53}
\ee
where Eq.(\ref{25}) and the near-classical approximation have again been
utilized.
\par
Interestingly, although not unexpectedly, 
$\Theta$  transforms just as would the stress tensor of a 
 conformal
field theory \cite{FMS}. This outcome can, at least conjecturally, be viewed
as a manifestation of the dS/CFT correspondence \cite{STR2,STR3}.
That is, in analogy to the 
AdS$_2$/CFT$_1$  correspondence \cite{STR4}, the asymptotic
symmetries of two-dimensional de Sitter gravity are
expected to be a subgroup of the 1+1-dimensional conformal
group. The third term in Eq.(\ref{53}) is the anticipated anomalous
derivative term, which is related to  the  central charge ($c$) 
of the associated Virasoro algebra \cite{FMS}. 
On this basis, it is possible to identify  $c/12=\eta_{0}$. 
\par
We can make the above observations more explicit by first
taking note of the following expression for the deformation algebra
\cite{BH}:
\be
\epsilon\DW\Theta =\{\epsilon\Theta,\omega\Theta\}_{DB},
\label{54}
\ee
where the subscript ``DB'' denotes the Dirac bracket formalism \cite{DIR}.
There are, however, subtle difficulties in directly applying 
this formalism to the case of a one-dimensional boundary \cite{CAD3}.
Nonetheless, Cadoni and Mignemi \cite{CAD4} have suggested
 the following circumvention: 
\be
\epsilon\DW\Theta=\{{\hat{\epsilon\Theta}},\omega\Theta\}_{DB}.
\label{55}
\ee
where the hat symbol signifies that, for a timelike (spacelike)
 boundary, this is
a time (position) averaged quantity.
\par
By way of Eqs.(\ref{53},\ref{55}) and the following Fourier
expansions:
\be
\epsilon=l\sum_{m}a_me^{imx/l},
\label{56}
\ee
\be
\omega=l\sum_{n}a_ne^{inx/l},
\label{57}
\ee
\be
\Theta={1\over l}\sum_{k}L_ke^{-ikx/l},
\label{58}
\ee
\be
{\hat{\epsilon\Theta}}={1\over l}\int^l_0dx \epsilon\Theta=
\sum_{m}a_m L_{m},
\label{59}
\ee
some straightforward calculation yields:
\be
\left[L_m,L_n\right]=(m-n)L_{m+n}+{12\eta_0\over 12}m^3\delta_{m+n,0}.
\label{60}
\ee
That is to say, the asymptotic symmetries of dS$_2$ are indeed realized
by a Virasoro algebra\footnote{The more conventional
form of the Virasoro algebra,
$\left[L_m,L_n\right]=(m-n)L_{m+n}+{c\over 12}m(m^2-1)\delta_{m+n,0}$,
can be obtained with a trivial shift: $L_0\rightarrow L_0+{1\over 2}
\eta_0$.} 
with a classical central charge of precisely:
\be
c=12\eta_0.
\label{666}
\ee
Notably, this is  the same value for the central
charge as found, via a different methodology,\footnote{Both
methods do, however, employ a canonical Hamiltonian framework.
See Section 1 for further discussion.}
 by Cadoni {\it et al} 
\cite{CAD2}.

\section{Cardy Entropy}

By way of \cite{CAD2}, we can already anticipate that the central
charge of Eq.(\ref{666}) will yield a statistical entropy in
agreement with the classical thermodynamic value (\ref{10}). 
However,  this 
coincidence provides
an important consistency check on the formalism, so let us
explicitly confirm that this is indeed the case.
\par
The calculation in question utilizes the well-known Cardy formula \cite{CAR},
which relates the density of states (i.e., statistical entropy)
of a CFT$_2$ to the associated Virasoro algebra. 
 This 
formula is obtained by
 way  of some formal manipulations 
 that exploit the modular and conformal invariance of the CFT$_2$ partition
function (for a recent derivation, see \cite{CARN1}). 
More specifically, one finds that:
\be
S_{Car}=2\pi\sqrt{{c\L_0\over 6}},
\label{61}
\ee
where  matters have been simplified with the (standard) assumptions
that $L_0>> c$ and  $L_0\sim 0$  for the  ground state.
\par
Before proceeding, let us point out that
the Cardy formula is, in fact, commonly employed
as an intermediate step in statistically motivated calculations of
black hole entropy (for example, \cite{STR}). Generally speaking,
one focuses on a relevant boundary 
(either a horizon or an asymptotic infinity) and then relates
the boundary dynamics to an effective Virasoro algebra.
(Such a relation is often possible because of
the high degree of symmetry of  black hole spacetimes.)
Although this methodology has enjoyed substantial
success, there are several  caveats  that suggest
the Cardy formula should be applied (in this manner)
with some degree of caution.
Most notably (see \cite{CARN2} for a thorough discussion):
{\it (i)} it is not apparent as to how  
an  asymptotic boundary theory can be sensitive
to the details of the interior geometry, whereas  it
is not evident that a black hole horizon
(which is an observer-dependent construct) can be regarded
as boundary at all, {\it (ii)} it has  not 
been rigorously  established that  {\it classical}
symmetries can be used to  determine the {\it quantum} properties of
a system, {\it (iii)} although such calculations are
naively statistical in nature, it is not at all clear
as to what   degrees of freedom are actually being
counted by  
the Cardy formula.  With regard to the last point,
it has, however, been argued that, for
the very special case of a two-dimensional (fundamental)  theory,
the  physical (black hole) degrees of freedom  
are truly being counted \cite{FUR}.
\par
Let us now consider the direct  application of Eq.(\ref{61})
to the current analysis. 
 Of course, $c$ is known (\ref{666}), but it is still necessary
to evaluate $L_0$ in terms of the bulk solution.
This is possible by way of Eq.(\ref{58}), which implies
the following  Fourier transform:
\be
L_n=\int^l_0dx \Theta e^{inx/l}.
\label{62}
\ee
Substituting for $\Theta=\Psi/\epsilon$ via Eq.(\ref{52}), we
have  for the classical solution (${\overline \rho}=0$):
\bea
L_0&=&l|M|\nonumber \\
&=& {\eta_0 a^2\over 2},
\label{63}
\eea
where Eq.(\ref{8}) has been used to obtain the lower line.
\par
It is now straightforward to evaluate Eq.(\ref{61}) for
the statistical entropy: 
\be
S_{Car}=2\pi\eta_0|a|,
\label{64}
\ee
in perfect agreement with the (classical) thermodynamic
entropy of Eq.(\ref{10}).

\section{Conclusion}

In summary, we have calculated the central charge in 
a  dS$_2$/CFT$_1$ context by  applying a methodology
that has been  advocated by Catelani, Vanzo and
Caldarelli \cite{VAN}. Our outcome substantiates
a recent, related calculation  by Cadoni {\it et al} \cite{CAD2}.
Although the two methods are both technically valid,
we have argued that, from a holographic perspective,
the current treatment is (perhaps) the aesthetically
preferable one.
\par
We went on to confirm that, given this value of
the central charge, Cardy's  statistical entropy
\cite{CAR} is in precise agreement with
the classical thermodynamic quantity. This
coincidence provides further support for
the conjectured  dS$_2$/CFT$_1$ \cite{NS,CAD2} correspondence.
This  relationship  may, of course, be a lower-dimensional realization
of a more general holographic duality.  
It is, however, a puzzlement that, given a dS bulk spacetime
of arbitrary dimensionality, the dually related CFT is
generally regarded as being a non-unitary theory \cite{STR2}. 
As an immediate consequence, such a theory can {\it not} be expected,
{\it a priori}, to conform with Cardy's entropic formula.
(Consult \cite{BMS} for further discussion.)
In all likelihood, this oddity will have to be better understood
before the dS/CFT duality can be ``promoted'' beyond
the conjectural level.

\section{Acknowledgments}

The author would like to thank V.P. Frolov for
helpful conversations.

\end{document}